\documentclass[page-classic,linenumbers]{epl2} 
\usepackage{amsmath}
\usepackage{latexsym}
\usepackage{amssymb}
\usepackage{graphicx}
\usepackage[colorlinks=true, citecolor=blue, urlcolor=blue]{hyperref}
\usepackage{float}
\usepackage{amsfonts}

\title{Measurement-device-independent randomness from local entangled states}

\author{Anubhav Chaturvedi$^1$, Manik Banik$^{2}$}

\institute{(1) Center for Computational Natural Sciences and Bio-informatics, IIIT-Hyderabad,\\ Gachibowli, Hyderabad 500032, India.\\
           (2) Optics \& Quantum Information Group, The Institute of Mathematical Sciences, C.I.T Campus, Tharamani, Chennai 600 113, India.
	  	   }

\shortauthor{A. Chaturvedi \& M. Banik}


\abstract{Nonlocal correlations are useful for device independent (DI) randomness certification [\href{http://www.nature.com/nature/journal/v464/n7291/full/nature09008.html} {Nature (London) {\bf 464}, 1021 (2010)}]. The advantage of this DI protocol over the conventional quantum protocol is that randomness can be certified even when experimental apparatuses are not trusted. 
Quantum entanglement is the necessary physical source for the nonlocal correlation required for such DI task. However, nonlocality and entanglement are  distinct concepts. There exist entangled states which produce no nonlocal correlation and hence are not useful for the DI randomness certification task. Here we introduce the measurement-device-independent randomness certification task where one has trusted quantum state preparation device but the mesurement devices are completely unspecified. Interestingly we show that there exist entangled states, with local description, that are useful resource in such task which otherwise are useless in corresponding DI scenario.}

\begin{document}

\maketitle
\section{Introduction}
Randomness is a valuable resource for various important tasks ranging from cryptographic applications \cite{Crypto} to numerical simulations such as \emph{Monte Carlo} method \cite{Monte}. Algorithmic information theory shows that true randomness cannot exist from a mathematical point of view \cite{Chaitin,Knuth}. Thus generation of randomness must be based on unpredictability of physical phenomena so that the random nature is guaranteed by the laws of physics. Classical physics being fundamentally deterministic in nature cannot guarantee such randomness \cite{Butterfield}. On the other hand though the outcomes of measurement performed on quantum system are intrinsically random (due to Born rule) \cite{Born,Neumann}, real-life implementation of such randomness generation procedures \cite{Jennewein,Stefanov,Atsushi} demand idealized modeling and detailed knowledge about the internal working process of the devices used for generating randomness. To overcome this issue, nonlocality based \cite{Ekert,Barrett_1,Masanes} and device independent (DI) technique \cite{Mayers,Acin_1,Colbeck,Pironio_1} has been applied for generating randomness. In Ref.\cite{Pironio_2}, Pironio \emph{et al.} have shown that correlation obtained from entangled quantum particles can be used to certify the presence of genuine randomness and they have designed cryptographically secure random number generator which does not require any assumption on the internal working of the devices. The key point is that randomness in the outcomes of measurements performed on the separated parts of the  entangled quantum systems can be certified in DI way if the correlation obtained from the entangled state violates a Bell inequality (BI). It is well known that nonlocality \cite{Brunner} and entanglement \cite{Horodecki} are two distinct concepts. Not all entangled states violate BI, rather there exists entangled states for which measurement statistics can be simulated locally \cite{Werner}. Therefore, such \emph{local} entangled states are not useful resource for DI randomness certification. In this work we first introduce the concept of measurement-device-independent (MDI) randomness certification protocol, where the quantum state preparation device behave quantum mechanically but the measurement device is completely untrusted. In such scenario we show that class of \emph{local} entangled states become useful resource for randomness certification task which otherwise are not useful for the corresponding DI scenario. 

The concept of MDI information processing scenario has been  independently introduced in Ref.\cite{Braunstein} and Ref.\cite{Lo}, where the authors have presented the idea of MDI-quantum key distribution (MDI-QKD) protocol. The important benefit of the MDI protocol over the conventional quantum one is that it requires no trust in the measurement device and hence comes the name. But, in contrast to DI protocols the MDI protocols require almost perfect state preparation device. 
Recently, Branciard \emph{et al.} have introduced another interesting protocol in MDI scenario. They have shown that presence of entanglement can be demonstrated in MDI way \cite{Branciard}. To arrive at their conclusion Branciard \emph{et al.} have used a recent result of Buscemi, which shows that all entangled states provide an advantage over the separable states for some a \emph{semi quantum} game \cite{Buscemi}.

In this work we first introduce the MDI randomness certification task. We then show that entangled states which are not useful for DI randomness certification turn out to be useful resource for the corresponding MDI scenario. More precisely we consider the two-qubit entangled Werner states $\varrho^v=v|\psi^-\rangle\langle\psi^-|+(1-v)\frac{\mathbb{I}}{2}\otimes \frac{\mathbb{I}}{2}$. It is known that Werner states with visibility parameter $v>1/3$ are entangled and a subclass of these states (states with $v>1/\sqrt{2}$) violates BI and hence are useful for DI randomness certification. On the other hand Werner states with $v\le 1/2$ and $v\le 5/12$ have local description for projective measurement and positive operator valued measurement (POVM), respectively \cite{Werner}, and thus cannot be useful for DI randomness certification. Interestingly, we show that all these entangled Werner states are useful for MDI randomness certification.

\section{Bell scenario and  DI randomness}\label{sec2}

A bipartite Bell scenario with $m$ different measurements per subsystem, each measurement having $d$ possible results, is characterized by the joint probabilities $P_{AB|XY}=\{p(ab|xy)\}$, with measurement results denoted by $a,b\in \{1,2,...,d\}$ and measurements denoted by $x,y\in \{1,2,...,m\}$. The quantum distribution $P_{AB|XY}^Q$ is of the form
\begin{equation}\label{q}
p(ab|xy)=\mbox{Tr} [M_{a|x}\otimes M_{b|y} \rho]
\end{equation}
where $\rho$ is a quantum state (density operator) in some tensor product Hilbert space $\mathcal{H}_A\otimes \mathcal{H}_B$ and $\{M_{a|x}~|~M_{a|x}\ge 0~\forall a;~\sum_{a}M_{a|x}=\mathbb{I}_{\mathcal{H}_A}\}$, $\{M_{b|y}~|~M_{b|y}\ge 0~\forall b;~\sum_{b}M_{b|y}=\mathbb{I}_{\mathcal{H}_B}\}$ are positive operator valued measures (POVMs) \cite{Nielson}. The set of quantum statistics $P_{AB|XY}^Q$ is referred to as $Q$. A Bell expression $I =\sum_{abxy}c_{abxy}p(ab|xy)$ is a linear combination of the probabilities specified by the coefficients $\{c_{abxy}\}$ \cite{Bell}. Correlations which can be expressed as $P(ab|xy)=\int_{\lambda}d\lambda\rho(\lambda)P(a|x,\lambda)P(b|y,\lambda)$ with $\lambda$ being the shared random variable, admit {\em local realistic} description and satisfy the condition $I\le I_L$, where $I_L$ is called the local bound of the BI. Interestingly, there exists entangled quantum states which violate BI and correlations obtain from these states can not be explained in local realistic form. Such correlations are called nonlocal correlations. However, there exists correlations which are more nonlocal than quantum correlation but compatible with relativistic causality or no signaling (NS) principle. The well known Popesku-Rohilick (PR) correlation \cite{Popescu} is an example of this type. If the collections of local, quantum and NS correlations are denoted as $\mathcal{P}^L$, $\mathcal{P}^Q$ and $\mathcal{P}^{NS}$, respectively, then the following strict set inclusion relations hold: $\mathcal{P}^L\subset \mathcal{P}^Q\subset \mathcal{P}^{NS}$ (see \cite{Brunner} for a review on Bell's nonlocality). Note that BI is derived under conjunction of the assumptions called \emph{reality} and \emph{locality} (along with \emph{measurement independence}). Violation of BI by quantum correlations implies that quantum mechanics is not reconcilable with these assumptions. As these assumptions refer to properties of a ontological (hidden-variable) model \cite{Rudolph}, thus from the observed BI violation it is impossible to conclude which one of these assumptions is violated. Interestingly, the BI can be derived under two operational assumptions, namely, \emph{predictability} and \emph{signal locality} \cite{Cavalcanti}. As the operational assumption of signal locality is an empirically testable (and well-tested) consequence of relativity, thus BI violation implies that events are unpredictable. This alternative derivation of BI from operational assumptions plays important role in the practical question of randomness certification even when the experimental devices are not trusted. 

In DI randomness certification scenario one (Say Alice) has a private place which is completely inaccessible from the outside i.e., no illegitimate system may enter in this place. From a cryptographic point of view assumption of such private place is admissible. Alice  chooses classical inputs $x\in X$ and $y\in Y$ with probability distributions $\mathcal{P}_X(x)$ and $\mathcal{P}_Y(y)$, respectively, and sends them to two measurement devices ($\mathcal{MD}1$ and $\mathcal{MD}2$ respectively) through some secure classical communication channels. The inputs prescribe the  measurement devices to perform some POVM $\{M_{a|x}~|~M_{a|x}\ge 0~\forall a;~\sum_{a}M_{a|x}=\mathbb{I}_{\mathcal{H}_A}\}$ and $\{M_{b|y}~|~M_{b|y}\ge 0~\forall b;~\sum_{b}M_{b|y}=\mathbb{I}_{\mathcal{H}_B}\}$ on some quantum state $\rho$, shared between the two devices. Once the inputs are received, no classical communication between the measurement devices $\mathcal{MD}1$ and $\mathcal{MD}2$ is allowed. Alice collects the input-output statistics $P(AB|XY)=\{p(ab|xy)\}$. 
\begin{figure}[b]
\centering
\includegraphics[height=4cm,width=6cm]{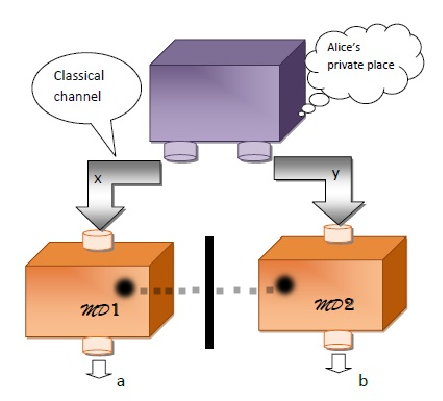}
\caption{(Colour on-line) Setup for DI randomness certification. Classical inputs $x,~y$ are sent from Alice's private place to the measurement devices $\mathcal{MD}1$ and $\mathcal{MD}2$, respectively, through secure classical channels. The black dots denote the bipartite quantum state $\rho$ shared between the two measurement devices. Classical communication is not allowed between two measurement devices.}\label{fig1}
\end{figure}
Since no communication between two measurement devices is allowed (i.e signal locality assumption is satisfied) hence BI violation implies that operational statistics must be unpredictable. Therefore randomness can be certified against an Eavesdropper with control of specifying the details of the experimental device. The setup for DI randomness certification is depicted in Fig.\ref{fig1}.

The amount of randomness associated with the measurement outcome is quantified by guessing probability $G(x,y,\mathcal{K})= \max_{a,b}p(ab|xy,\mathcal{K})$ \cite{Acin_2} of a malicious Eavesdropper who prepares the experimental devices. Here $p(ab|xy,\mathcal{K})$ are the joint outcome probabilities and $\mathcal{K}$ denotes the shared resources between the two spatially separated system. If the Eavesdropper is restricted by quantum theory then she prepares $\mathcal{K}$ as any bipartite quantum state. On the other hand, if she is restricted only by no signaling (NS) principle then $\mathcal{K}$ can be any correlation satisfying NS principle. The quantity G corresponds to the Eave s probability to guess correctly the outcomepair $(a, b)$, since the best guess is simply to output the most probable pair. The guessing probability can be expressed in bits and is then known as the min-entropy, $H_{\infty}(x,y,\mathcal{K})=-\log_2G(x,y,\mathcal{K})$ \cite{Koenig}. In \cite{Pironio_2}, Pironio \emph{et al.} have shown that whenever a bipartite input-output probability distributions violates BI there is nonzero min-entropy associated with the outputs. To obtain the minimum randomness in quantum theory one has to perform the following optimization problem: 
\begin{eqnarray}\label{rand_quantum}
p^*_q(ab|xy)&=&~~~~~~\mbox{max}~~~~~~p(ab|xy)\nonumber\\
&&\mbox{subject~to}~~\sum_{abxy}c_{abxy}p(ab|xy)=I\nonumber\\
&& p(ab|xy)\mbox{~is~quantum},
\end{eqnarray}
where the last condition ensures that the obtained correlation is of the form Eq.(\ref{q}). Adapting a straightforward way of technique for approximating the set of quantum correlations using a semi-definite-programs (SDP) as introduced
in \cite{Navascues}, one can efficiently lower bound min-entropy obtainable from a quantum correlation. The minimum random bits obtained in quantum theory corresponding to BI violation $I$ is thus $H_\infty(AB|XY)=-\log_2\max_{ab}p_q^*(ab|xy)$. One may, however, be interested in the amount of randomness obtained in NS theory; which mean that instead of the quantum state any correlation satisfying NS condition is allowed to share between the measurement devices (see \cite{Pironio_2} for NS analysis).

\section{Semi-quantum nonlocal game scenario}\label{sec3}
Recently, Buschemi generalizes the standard Bell game scenario into semi quantum scenario \cite{Buscemi}. In this case Alice chooses classical inputs $x\in X$ and $y\in Y$ with probability distributions $\mathcal{P}_X(x)$ and $\mathcal{P}_Y(y)$, respectively. But, instead of sending these classical inputs to the measurement devices she encodes the information of these inputs into sets of quantum states $\{|\phi^x\rangle_{\alpha'}\}_{x\in X}$ and $\{|\psi^y\rangle_{\beta'}\}_{y\in Y}$, chosen from Hilbert spaces $\mathcal{H}_{\alpha'}$ and $\mathcal{H}_{\beta'}$, respectively. The quantum states $|\phi^x\rangle$ and $|\psi^y\rangle$ are then send to the measurement devices $\mathcal{MD}1$ and $\mathcal{MD}2$, respectively, through quantum channels. Given these quantum states the respective measurement device $\mathcal{MD}1$ and $\mathcal{MD}2$ produce outcomes $a$  and $b$, respectively, by performing POVMs on the composite system i.e. the system obtained from Alice and the part of a bipartite state $\rho_{\alpha\beta}$, shared between the two measurement devices $\mathcal{MD}1$ and $\mathcal{MD}2$. The output probability  is 
\begin{eqnarray}\label{defQ}
p_{\rho_{\alpha\beta}}(ab||\phi^x\rangle_{\alpha'},|\psi^y\rangle_{\beta'})
=\mbox{tr}[(\mathcal{M}^{\alpha'\alpha}_a\otimes \mathcal{M}^{\beta\beta'}_b)\nonumber\\
(|\phi^x\rangle_{\alpha'}\langle\phi^x|\otimes \rho_{\alpha\beta}\otimes |\psi^y\rangle_{\beta'}\langle\phi^y|)],
\end{eqnarray}  
where $\mathcal{M}^{\alpha'\alpha}_a$ ($\mathcal{M}^{\beta\beta'}_b$) is the element of the POVM performed on the composite system $\mathcal{H}_{\alpha'}\otimes \mathcal{H}_{\alpha}$ ($\mathcal{H}_{\beta}\otimes \mathcal{H}_{\beta'}$) to produce the outcomes $a$ and $b$. Expression of Eq.(\ref{defQ}) can also be written as,
\begin{eqnarray}\label{defQ1}
p_{\rho_{\alpha\beta}}(ab||\phi^x\rangle_{\alpha'},|\psi^y\rangle_{\beta'})
=\mbox{Tr} [M_{a||\phi^x\rangle_{\alpha'}}\otimes M_{b||\psi^y\rangle_{\beta'}} \rho_{\alpha\beta}],
\end{eqnarray}
where the operators $M_{a||\phi^x\rangle_{\alpha'}}=$ and $M_{b||\psi^y\rangle_{\beta'}}$ describe Alice and Bob’s effective POVMs acting on $\rho_{\alpha\beta}$ given $|\phi^x\rangle_{\alpha'},|\psi^y\rangle_{\beta'}$. We shall refer to the set of quantum probabilities of the form of Eq.(\ref{defQ1}) as $Q$.

In this generalized framework Buscemi proved that if the shared state between the measurement devices $\mathcal{MD}1$ and $\mathcal{MD}2$ is entangled one then Alice can choose the input quantum states in such way that the produced correlation cannot be achieved by local operation and shared randomness (LOSR). Later it has been shown that in this scenario any entangled state can generate correlations that cannot be simulated by local operation and classical correlation (LOCC) even if there is no restriction on the amount of classical communication \cite{Rosset}, but that such correlations can be simulated if the distribution of the shared variables depends on the input quantum states i.e., it the measurement independence assumptions have been reduced \cite{Banik}. Using these semi quantum game framework, in the following, we explicitly show that all two-qubit entangled Werner states are useful for MDI randomness certification. 

We consider the following particular semi-quantum game. The input quantum states are chosen from a regular tetrahedron on the Bloch sphere i.e., 
\begin{equation}\label{qstate}
|\phi^x\rangle\langle\phi^x|=\frac{\mathbb{I}+\vec{v}_x.\vec{\sigma}}{2},
~~|\psi^y\rangle\langle\psi^y|=\frac{\mathbb{I}+\vec{v}_y.\vec{\sigma}}{2},
\end{equation}     
for $x,y=1,..,4$ we have $\vec{v}_1=\frac{(1,1,1)}{\sqrt{3}}$, $\vec{v}_2=\frac{(1,-1,-1)}{\sqrt{3}}$, $\vec{v}_3=\frac{(-1,1,-1)}{\sqrt{3}}$ and $\vec{v}_4=\frac{(1,-1,-1)}{\sqrt{3}}$; and $\vec{\sigma}=(\sigma_1,\sigma_2,\sigma_3)$ with $\sigma_i$ ($i=1,2,3$) being the Pauli matrices. The POVM $\{\mathcal{M}^{\alpha'\alpha}_a\}_{a\in \{0,1\}}$ is given by 
\begin{equation}
\mathcal{M}^{\alpha'\alpha}_1=|\phi^+\rangle\langle\phi^+|,~~\mathcal{M}^{\alpha'\alpha}_0=\mathbb{I}-|\phi^+\rangle\langle\phi^+|,
\end{equation}
\begin{figure}[t]
\centering
\includegraphics[height=4cm,width=6cm]{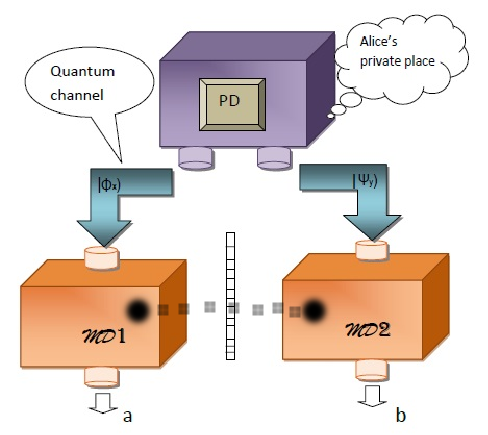}
\caption{(Colour on-line) Setup for MDI randomness certification. Alice has perfect state preparation device (PD) at her private place. Quantum states $|\phi^x\rangle_{\alpha'}$ and $|\psi^y\rangle_{\beta'}$ are sent from Alice's private place to the measurement devices $\mathcal{MD}1$ and $\mathcal{MD}2$, respectively, through secure quantum channels. Black dots are the quantum state $\rho_{\alpha\beta}$ shared between two devices. Classical communication is allowed between two measurement devices but no quantum state transfer is allowed.}\label{fig2}
\end{figure}
where $|\phi^+\rangle=\frac{|00\rangle+|11\rangle}{\sqrt{2}}$. Same POVM is considered at Bob's end $\{\mathcal{M}^{\beta\beta'}_b\}_{b\in \{0,1\}}$.The probability distribution admitted when $\rho_{\alpha\beta}$ is a singlet state is,
\begin{equation}\label{p1}
p(a,b||\phi^x\rangle,|\psi^y\rangle)=    \begin{cases}
\frac{2-(a+b)}{4}, & \text{if}\ x=y\\
\frac{7-5a-5b+4ab}{12}, & \text{if}\ x \neq y\\
\end{cases}
\end{equation}
Notice The probability distribution admitted when $\rho_{\alpha\beta}$ is $\frac{\mathbb{I}}{4}$,
\begin{equation}\label{p2}
p(a,b||\phi^x\rangle,|\psi^y\rangle)=    \begin{cases}
\frac{9}{16}, & \text{if}\ a=0 \text{and}\ b=0\\
\frac{3}{16}, & \text{if}\ (a \oplus b=1) \\
\frac{1}{16}, & \text{if}\ a=1 \text{and}\ b=1 \\
\end{cases}
\end{equation}
for all $x,y$. The Werner state is a classical mixture of these two states and hence the prbability distribution.

 It is known that $W=\frac{\mathbb{I}}{2}-|\psi^-\rangle\langle\psi^-|$ is an entanglement witness for the two-qubit Werner state $\varrho^v$ \cite{Toth}. For Werner state $\varrho^v$, $\mbox{tr}[\varrho^vW]=\frac{1-3v}{4}$, which is negative for $v>\frac{1}{3}$ and $\mbox{tr}[\rho W]>0$ for any separable state $\rho$. From this entanglement witness operator Branciard \emph{et al.} have constructed the following MDI-entanglement witness \cite{Branciard}:
\begin{equation}\label{mdi-ew}
I(P)=\frac{5}{8}\sum_{x= y}p(1,1||\phi^x\rangle,|\psi^y\rangle)-\frac{1}{8}\sum_{x\ne y}p(1,1||\phi^x\rangle,|\psi^y\rangle).
\end{equation}     
Here $P$ denotes the probability distribution $\{p(a,b||\phi^x\rangle,|\psi^y\rangle)|a,b=0,1;x,y=1,..,4\}$. For the Werner states the above expression becomes $I(P_{\varrho^v})=\frac{1-3v}{16}$, which is negative for $v>\frac{1}{3}$. For any separable state $\rho$, $I(P_{\rho})=0$, as separable states are the end points of the semi-quantum game relation `$\succcurlyeq_{sq}$' defined in \cite{Buscemi}.

\section{MDI randomness certification}\label{sec4}
We are now in the position to show that any two-qubit entangled Werner states can certify the presence of randomness when the measurement apparatuses are not trusted. The set up for MDI randomness certification is depicted in Fig.\ref{fig2}. Here, in contrast to the DI randomness certification scenario (Fig.\ref{fig1}), Alice has a perfect state preparation device at her private place. The quantum states, chosen from the set described in Eq.(\ref{qstate}) are prepared by Alice and are sent to measurement devices $\mathcal{MD}1$ and $\mathcal{MD}2$ through quantum channels. No leakage of the information about the classical index $x$ (or $y$) is allowed. In DI scenario, after sending the classical index $x$ and $y$ to the respective measurement devices no classical communication is allowed between the measurement devices. In this case no such restriction is required. But after receiving the quantum states from Alice any kind of quantum state transfer is prohibited between the two measurement devices. When the quantum states reach to the measurement devices, both the devices produce classical outcomes $a,b\in \{1,0\}$. Alice collects the input-output statistics and tests whether the the collected data satify certain conditions.

{\bf Results}:
To find the minimum randomness associated with the probability distribution $P=\{p(ab|xy)\}$ one has to solve the following optimization problem,

\begin{eqnarray}\label{mdi_rand_quantum}
p^*(ab|xy)&=&~~~~~~\mbox{max}~~~~~~p(ab|xy)\nonumber\\
&&\mbox{subject~to}~~I(P)=\frac{1-3v}{16}\nonumber\\
&& p(ab|xy)\in Q,
\end{eqnarray}
where $I(P)$ is the expression of Eq.(\ref{mdi-ew}).
The minimum random bits obtained in quantum theory corresponding to Werner state visibility parameter $v$ is thus $H_\infty(AB|XY)=-\log_2\max_{ab}p_q^*(ab|xy)$. While the optimization problem (\ref{mdi_rand_quantum}) is computationally tough, one can solve for a relaxed condition $p(ab|xy)\in Q_{1+AB}$ using SDP. Alternatively $p(ab|xy)\in NS$ can be used to quantify minimum random bits obtained from no-signaling principle.
Our results point out that there is zero min-entropy  against a $Q_{1+AB}$ and no-signaling (see Appendix). 
Changing the visibility parameter $v$ given each free runs of the protocol corresponds to movement on the line joining (\ref{p1}) and (\ref{p2}) in the probability distribution space (two party quadruple inputs binary output). Hence we look for characteristics of (\ref{p1}) and (\ref{p2}) that guarantee randomness. 

\emph{Additional conditions on statistics:} 
As the protocol used above is same up-to relabeling for outputs, Eq.(\ref{mdi-ew}) in general can be written as,
\begin{equation}\label{mdi-ew1}
I(P)=\frac{5}{8}\sum_{x= y}p(i,j||\phi^x\rangle,|\psi^y\rangle)-\frac{1}{8}\sum_{x\ne y}p(i,j||\phi^x\rangle,|\psi^y\rangle)
\end{equation}
where $i,j\in\{0,1\}$. For $i=j$, following two conditions (should hold simultaneously) are sufficient for guaranteeing randomness (positive min-entropy) associated with the distribution $P$ for the parameter ranges $v\in(\frac{1}{3},1]$ under no-signaling and $Q_{1+AB}$.\\
Condition (I): \begin{equation}
P(0,1|l,l)=P(0,1|m,m),~   \forall~ l,m\in \{1,2,3,4\}, 
\end{equation}
i.e. when Alice and Bob have the same input the probability of obtaining outcomes $a=0$ and $b=1$ should be the same.\\
Condition (II): \begin{equation}
P(1,0|l,l)=P(1,0|m,m),~  \forall~ l,m\in \{1,2,3,4\}, 
\end{equation}
i.e. when Alice and Bob have the same input the probability of obtaining outcomes $a=1$ and $b=0$ should be the same.

\begin{figure}[t]
\centering
\includegraphics[height=4cm,width=7cm]{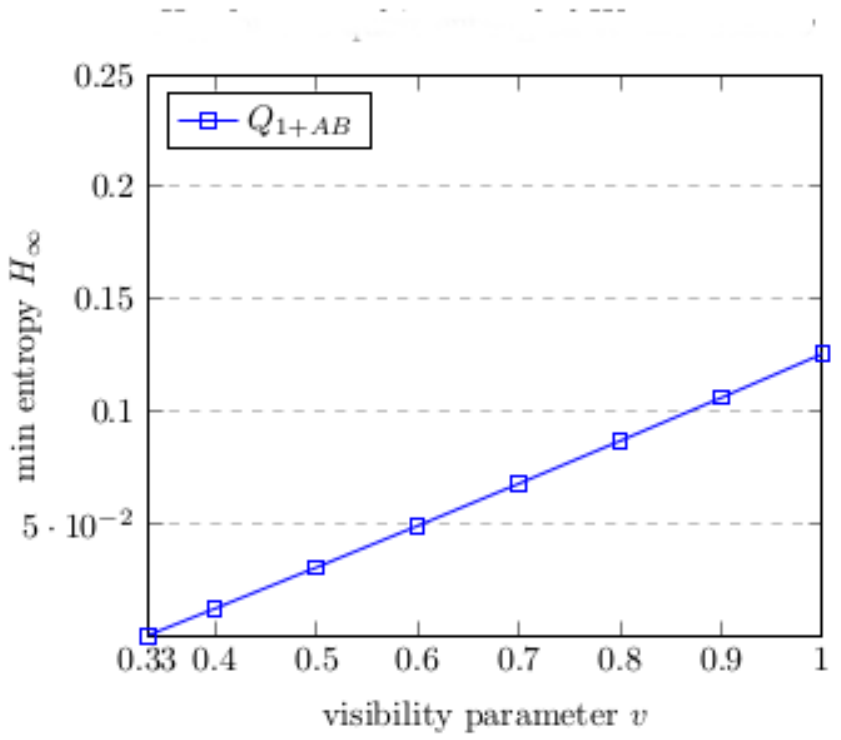}
\caption{(Colour on-line) The min-entropy statistics obtained by solving the optimization problem (\ref{mdi_rand_quantum}) using $Q_{1+AB}$ level of NPA hierarchy against visibility parameter $v$ under conditions (I) and (II). The min entropy under NS condition is same as obtained under $Q_{1+AB}$.}\label{plot1}
\end{figure}

After performing the optmization of Eq.(\ref{mdi_rand_quantum}) with the aditional conditions (I) and (II) the min-entropy is plotted in Fig. \ref{plot1}. 
However for the case when $i\neq j$ the following two conditions produce the same statistics as in Fig. \ref{plot1}.
Condition (III): \begin{equation}
P(0,0|l,l)=P(0,0|m,m),~   \forall~ l,m\in \{1,2,3,4\}, 
\end{equation}
i.e. when Alice and Bob have the same input the probability of obtaining outcomes $a=0$ and $b=0$ should be the same.\\
Condition (IV): \begin{equation}
P(1,1|l,l)=P(1,1|m,m),~  \forall~ l,m\in \{1,2,3,4\}, 
\end{equation}
i.e. when Alice and Bob have the same input the probability of obtaining outcomes $a=1$ and $b=1$ should be the same.

It is important to note that positive min entropy is obtain for $I(P)<0$, the condition which is satisfied by any two qubit entangled Werner states. Moreover no seperable state satifies this condition hence no cheating strategy is possible by sharing seperable correlations. Two qubits entangled Werner class of states also satified the additional conditions and hence they are useful for MDI min-entropy (randomness) certification. Also we obtain that the min entopy graph of the optimazation problem (\ref{mdi_rand_quantum}) (along with the additional conditions) in NS scenario (i.e. $p(ab|xy)\in NS$) is same as $Q_{1+AB}$  plotted in Fig.\ref{plot1}.

\section{Discussion}\label{sec5}
Specifying various device independent protocols based on the study of quantum nonlocality has importance practical implications. Various such potocols has been reported \cite{DI1,DI2,DI3,DI4,DI5} some with experimental realization. Among these one of the very interesting is DI randomness certification and generation. Violation of Bell- inequality guarantees randomness even from uncharacterised experimental devices. Nevertheless, the practical implementation of such protocols is extremely challenging as it requires the genuine violation of Bell’s inequality \cite{DI-Ex}. So different variant of randomness certification protocol has been reported which requires some assumptions on the devices. As for example Ref.\cite{Lunghi} degine a practical self testing QRNG protocol which requires some knowledge about the dimension of the quantum systems used in the protocol. However in all such DI or semi DI independent prtocols only those entangled states are useful that exhibit nonlocality. Hoever there exist entangled states which are local even under (nonsequential) generalized measurement.

Here we introduce the MDI randomness certification protocol which requires trusted quantum state preparation device but the measurement device is completely unspecified i.e. it can be supplied even by eavesdropper. In this scenario we show that some \emph{local} entangled states become useful in the task which othwise were useless in the corresponding DI secnario. One practicle advantage of our protocol over the DI or semi DI protocol is that in the DI scenario (see Fig.\ref{fig1}) any particle transfer or field interaction with the potentiality of sending classical communication between the two measurement devices need to be bolcked. But in our MDI scenarion this requirment is relaxed. One need not to bother about the classical communications between these devices but of course no quantum state tranfer between the devices is allowed.

Our works motivates further research. First of all note that we have considered single shot scenario and the protocol presented here is not optimal one. It is interesting to find the optimal protocol and then compare its rate with the DI protocol. On the other, it is also interesting to study where all entangled state are useful for the MDI randomness certification task. In Ref.\cite{Koh} the authors have shown that relaxation of `measurement independence' assumption in Bell's theorem potentially enhance the adversary's capabilities in the task of randomness expansion. In Ref.\cite{Banik} one of the author of this letter has shown that correlations achieved in semi-quantum nonlocal game scenario can be simulated by reducing `measurement independence'. In light of these two results it will be interesting to study the effect of reduced `measurement independence' in MDI randomness certification task.

\section{Acknowledgments}
MB likes to thank G. Kar for simulating discussions. Discussions with D. Rosset at ISI-Kolkata and comments of A. Ac\'{i}n in a private communication are gratefully acknowledged by MB. It is a great pleasure to thank T. Chakraborty for the help in improving the presentation of the manuscript. MB likes to acknowledge ISI (Kolkata), as part of this work is done there during his PhD.

\section{Appendix}
\begin{figure}[t]
\centering
\includegraphics[height=5cm,width=7cm]{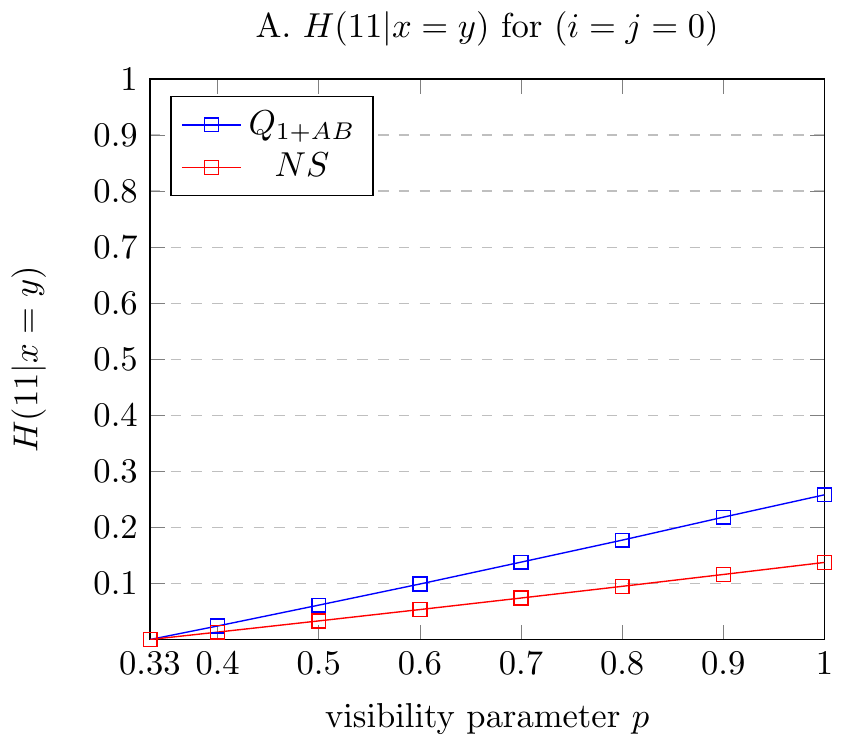}
\includegraphics[height=5cm,width=7cm]{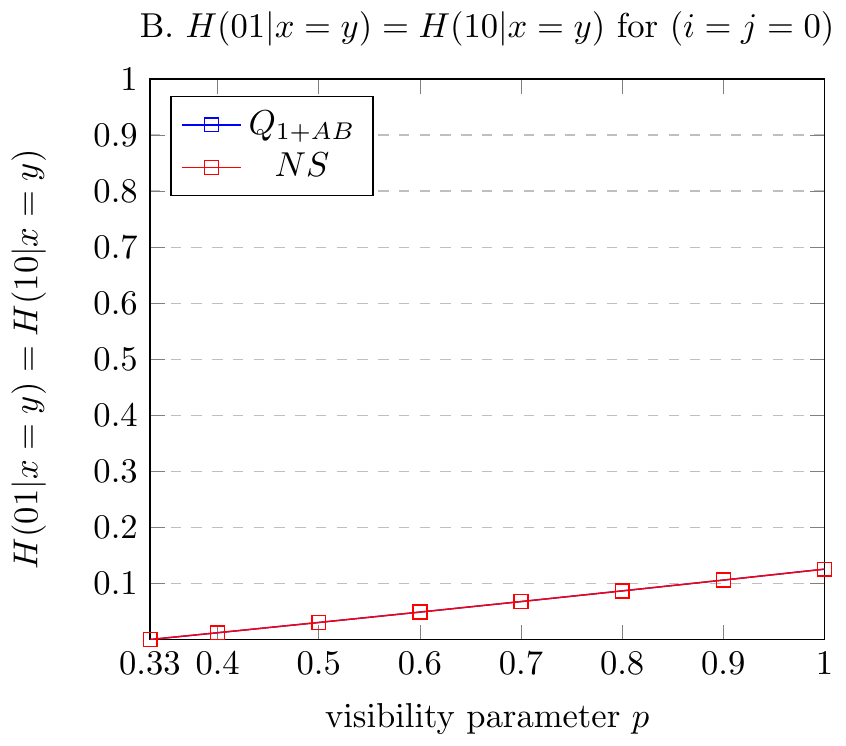}
\includegraphics[height=5cm,width=7cm]{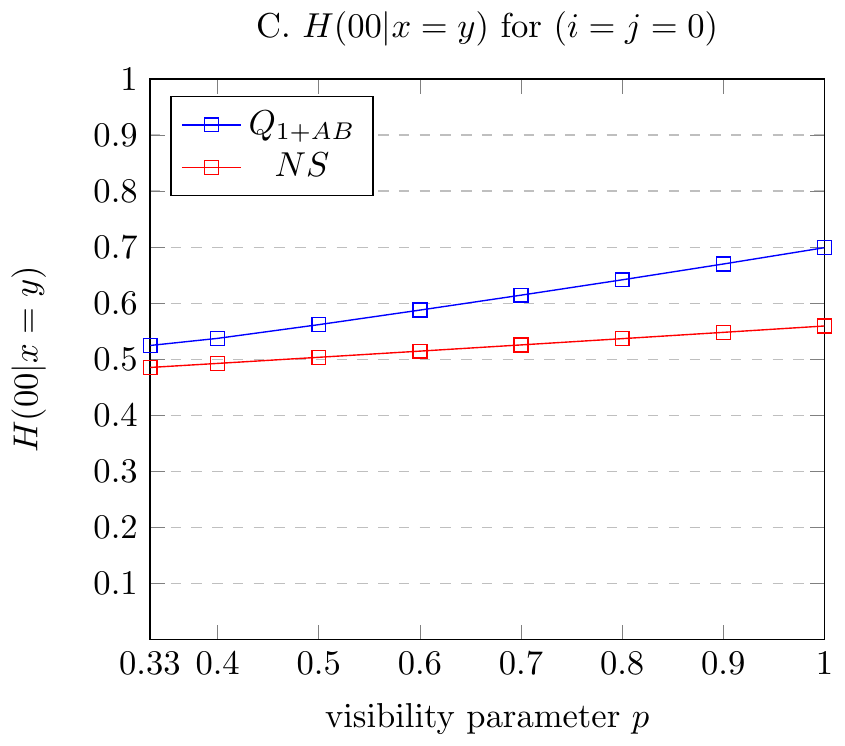}

\caption{The $H(ab|x=y)$ statistics obtained by solving the optimization problem (\ref{mdi_rand_quantum}) of the form using $Q_{1+AB}$ level of NPA hierarchy (blue) and no-signaling (red) against visibility parameter $v$ along with $i=j=0$ in (\ref{mdi-ew1}).}\label{plot2}
\end{figure}

\begin{figure}[t]
\centering
\includegraphics[height=5cm,width=7cm]{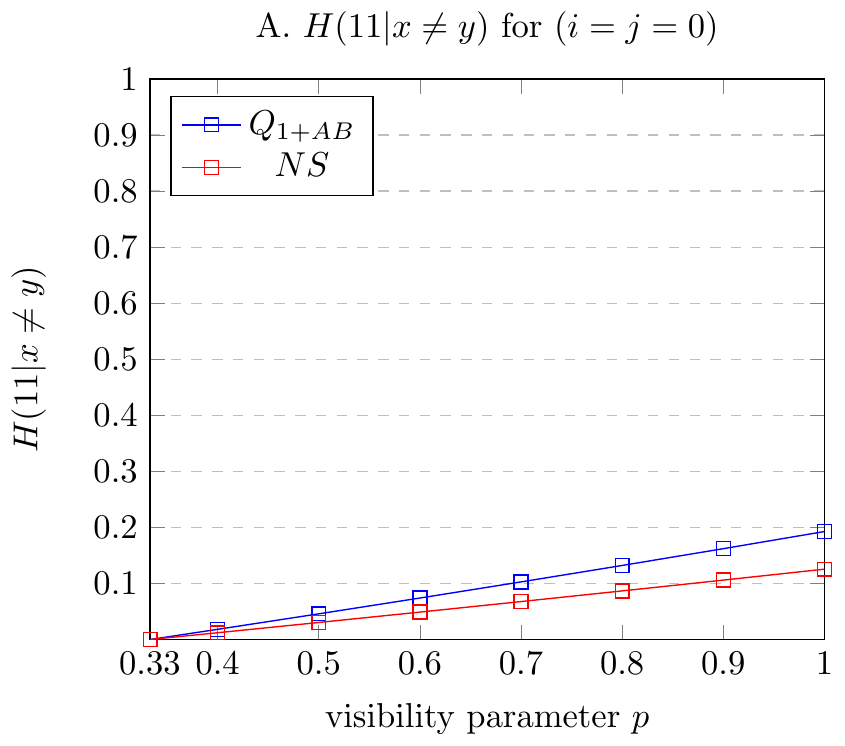}
\includegraphics[height=5cm,width=7cm]{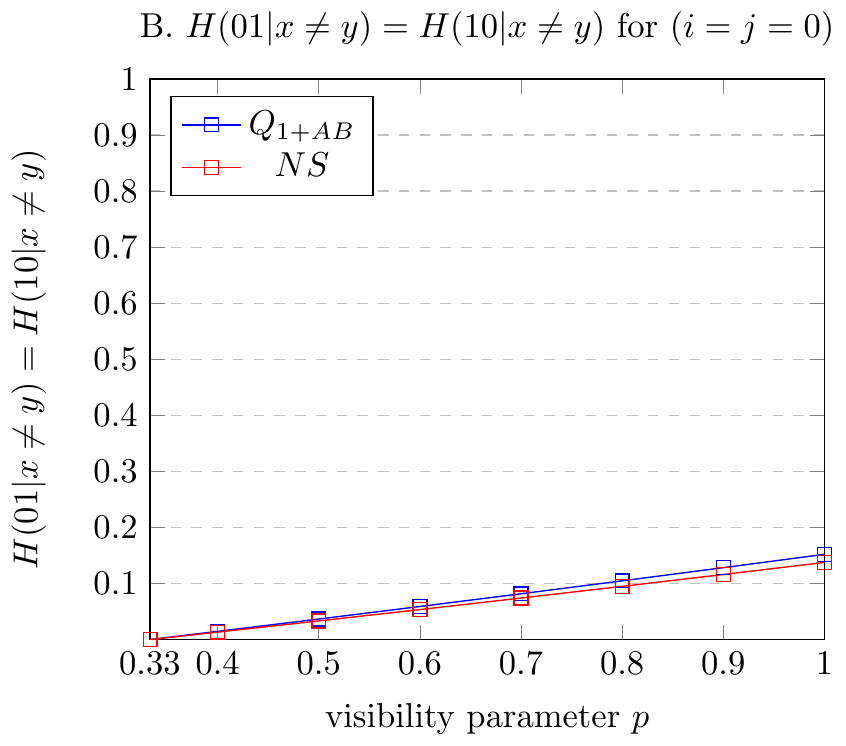}
\includegraphics[height=5cm,width=7cm]{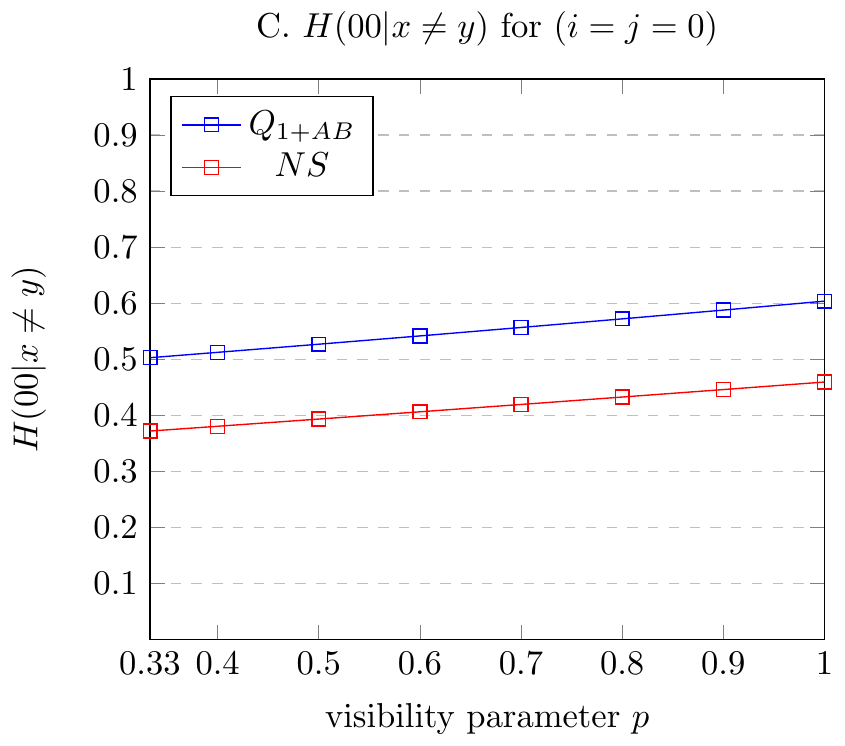}

\caption{The $H(ab|x=y)$ statistics obtained by solving the optimization problem (\ref{mdi_rand_quantum}) of the form using $Q_{1+AB}$ level of NPA hierarchy (blue) and no-signaling (red) against visibility parameter $v$ along with $i=j=0$ in (\ref{mdi-ew1}) and Conditions 1 and 2 hold.}\label{plot3}
\end{figure}

We use the perspective of the Eave's dropper to present the results. Eve prepares measurement apparatus for Alice. The optimization problem (\ref{mdi_rand_quantum}) can be seen as Eave's best strategy to increase the guessing probability  $p_q^*(ab|xy)$ of outcome $ab$ given inputs $xy$. The min-entropy,
\begin{equation}
H_\infty(AB|XY)=-\log_2\max_{ab}p_q^*(ab|xy).
\end{equation}
For all $v\in(1/3,1]$ without any extra condition Eave could always find $ab$ such that $H(ab|xy)$ is zero for both $Q_{1+AB}$ and no-signaling correlations which implies zero $H_\infty(AB|XY)$. However she gets positive $H(00|xy)$ when $x=y$ for some $v\in (1/3,1]$.
Under the Conditions (I) and (II) $H(ab|x=y)$ statistics are given in Fig. \ref{plot2} and $H(ab|x\neq y)$ statistics are given in Fig. \ref{plot3}. Notice best strategy for (no-signaling or $Q_{1+AB}$) Eave in both the cases ($x=y$ or $x\neq y$) is to maximize the guessing probability of $p(01|xy)$ or $p(10|xy)$. 

\end{document}